\documentclass{elsart}

\usepackage{epsfig,xspace}

\begin{document}

\begin{frontmatter}
\title{Issues and methods for CMB anisotropy data reduction}
\runtitle{Issues and methods for CMB anisotropy data reduction}
\author[CdF]{Jacques Delabrouille}
\address[CdF]{Physique Corpusculaire et Cosmologie, Coll\`ege de France 
\\
11, place Marcelin Berthelot, 75231 Paris cedex 05}
\runauthor{J. Delabrouille}
\begin{abstract}
Major issues and existing methods for the reduction of CMB anisotropy data
are reviewed. An emphasis is put on the importance of proper 
modelling of the data. It is suggested that the robustness of methods could 
be improved by taking into account the uncertainty of the model for 
finding optimal solutions. 
\end{abstract}

\begin{keyword}
methods: data analysis, methods: numerical, 
cosmology: cosmic microwave background
\PACS 98.80-k,98.70.Vc,95.75-z,95.75.Pq
\end{keyword}
\end{frontmatter}

\section{ Introduction } \label{sec:intro}

The importance of measuring anisotropies of the Cosmic Microwave 
Background (CMB) to constrain cosmological models is now well 
established. It has been demonstrated that the properties of these 
anisotropies depend drastically on the seeds for structure formation 
as well as on cosmological parameters describing the matter content, 
the geometry, and the evolution of our Universe 
\cite{hu-sugiyama96,jungman96}.

The complexity of CMB data taking into account imperfections in the 
observations, however, has led 
the CMB community into developing sophisticated data reduction methods.
In this review paper, I discuss the principal issues about CMB data analysis,
centering the discussion on map-making, as other aspects related to 
$C_{\ell}$ estimation and its interpretation are discussed 
by other authors in these proceedings.

\section{ Modelling the data for its reduction }

Data reduction methods involve:
\begin{itemize}
    \item writing down an accurate model of the problem (including 
    prior knowledge);
    \item deciding optimisation criteria;
    \item finding the optimal solution given the model and the 
    optimisation criteria;
    \item implementing the solution numerically.
\end{itemize}

The last two aspects have been discussed extensively in the 
litterature in the context of CMB anisotropy measurements
\cite{whb96,tegmark2-97}. Less 
emphasis has been put on the first two aspects so far, although they 
are certainly as important as the last two.

{\em modelling}

For most CMB experiments, the problem is modelled as trying 
to measure CMB temperature fluctuations $T_{p}$ 
of the sky in a number of sky pixels, indexed here by $p$, having
measurements in the form of a series of samples $m_{t}$ (in the
context of map-making, for instance, samples can typically be indexed 
by time). The simplest way of expressing the measurements $m_{t}$
as a function of the unknown $T_{p}$ is:

\begin{equation}
    m_{t} = A_{tp}T_{p} + n_{t}
    \label{eq:simple-model}
\end{equation}
here, $T_{p}$ and $m_{t}$ are one-dimensional 
vectors indexed by pixel numbers and by time. $A_{tp}$ is a pointing 
matrix telling how much of pixel $p$ is seen at time $t$. $n_{t}$ 
is the noise. Summation over repeated indices is assumed.

If we have several timelines corresponding to several 
detectors, the measurement can be modelled as one single data vector by 
appending all data streams one after the other.
Similarly, when the sky emission, represented by 
temperature $T_{p} = \sum_{c} T_{p}^c$ is a superposition of emissions 
due to various astrophysical processes, the system can 
be written in the form of equation \ref{eq:simple-model} with $p$ 
indexing both pixel and component (instead of pixel only). If there 
are $N_{c}$ components (galactic dust, CMB, Sunyaev Zel'dovich 
emission,\ldots), the linear system becomes simply $N_{c}$ times as large 
(with $N_{c}$ times more unkowns).
Formulation \ref{eq:simple-model} 
is thus quite general (up to the constraint that the model is linear).

It is often assumed that the pointing matrix is perfectly well 
known, that the noise is Gaussian 
(and stationnary) and that its 
autocorrelation $N = \langle n\,n^t\rangle$ (the so-called noise 
covariance matrix) is known ($n^t$ is the transpose of vector $\vec 
n$).

{\em Optimisation criteria}

The ``Optimisation" of the processing requires
an optimisation criterium. For the construction of a pixellised map 
one wants typically to minimise the quadratic sum of the errors in 
all pixels:
\begin{equation}
    \chi^2 = \sum_{p} ({\widetilde T}_{p} - T_{p})^2
    \label{eq:chi2-pixels}
\end{equation}
where ${\widetilde T}_{p}$ denotes the estimator of the true pixel 
temperature $T_{p}$.
If the distributions for $n$ and $T$ are Gaussian and their 
autocorrelations $N$ and $S$ are known, minimising the $\chi^2$ of equation 
\ref{eq:chi2-pixels} leads to the Wiener solution:
\begin{equation}
    {\widetilde T} = [S^{-1} + A^tN^{-1}A]^{-1} A^tN^{-1} \, m
    \label{eq:wiener}
\end{equation}

Alternatively, assuming no prior knowledge on the sky signal,
one may decide to minimize a standard $\chi^2$ on the measurements:
\begin{equation}
    \chi^2 = (m-A{\widetilde T})^t \, N^{-1} \, (m-A{\widetilde T})
    \label{eq:chi2-measure}
\end{equation}
in which case the solution is the {\it COBE} method, i.e.:
\begin{equation}
    {\widetilde T} = [A^tN^{-1}A]^{-1} A^tN^{-1} \, m
    \label{eq:COBE}
\end{equation}

The Wiener method is the optimal linear method when all processes are 
Gaussian and when their spectra are known. Non-linear methods relying 
on Maximum Entropy, using Neural Networks, or linear methods using 
wavelets may perform better in a variety of cases 
\cite{hjlb98,baccigalupi00,cayon00}.

Both the Wiener and the {\it COBE} methods rely on a particular prior 
knowledge about the signal and the noise: the knowledge of 
their autocovariance.  The availability of such prior information is 
far from being obvious, as stressed by Ferreira and Jaffe 
\cite{ferreira-jaffe00}, even if it does not need to be perfect.  

The results of the data reduction depend to some extent on 
the reliability of this knowledge.
If we assume that the noise is white (uncorrelated), for instance, 
the {\it COBE} method reduces to simple averaging of measurements
in pixels:
\begin{equation}
    {\tilde T} = [A^tA]^{-1} A^t \, m
    \label{eq:simple-average}
\end{equation}
If the assumption is erroneous (i.e. the noise is not white), this 
map-making method is very far from being 
optimal~\cite{tegmark2-97,jd-destripe98}.

More generally, the results of the data reduction depend on the 
accuracy of the model. In the future, methods taking into account 
explicitely the confidence one has in the model should be implemented. 
This is the object of current research. I believe it will help finding the 
best solution, as well as quantifying the errors, making results more 
robust.

{\em Optimal solution}

Once the problem is properly modelled, finding the optimal 
mathematical solution to data reduction is generally not the most 
difficult task for the applications we are interested in. The 
assesment of errors in the case of non gaussian processes, however, 
is not simple. So far, linear methods with the assumption of 
Gaussianity have performed satisfactorily.

{\em Implementation}

The numerical implementation of even the simplest data reduction 
scheme is quite a formidable problem in the context of present and 
future CMB missions, as emphasised by Bond et al.
\cite{bcjk99}. Iterative methods to solve large systems can 
be employed in some cases \cite{whb96}, but implementing the optimal 
solution for a large and complex system as for the upcoming Planck 
mission is a hard task.

\section{ Main issues for data reduction and interpretation }
\label{sec:issues}

We now review and discuss specific examples which constitute
some of the major issues about CMB observation and data analysis:

\begin{enumerate}
    \item astrophysical foregrounds
    \item noise and slow drifts (the so-called 1/f noise)
    \item imperfections of instrument properties and of their knowledge
\end{enumerate}

{\em Astrophysical components}

CMB anisotropy experiments observe the sky at frequencies ranging from 
few GHz to few hundreds of GHz, at centimeter to millimeter 
wavelengths.  The observed emission is not only due to CMB 
anisotropies, as other astrophysical processes also emit in this 
frequency range \cite{fb-rg99}.  The major known contributors are 
three galactic processes (thermal galactic dust emission, 
bremsstrahlung emission of free electrons -- the so-called free-free 
emission, synchrotron emission) and extragalactic sources (IR and 
radio galaxies, and hot ionised gaz through the SZ effect, principally 
in clusters of galaxies). In data processing, one needs to separate 
the contribution of different sources. This aspect will be discussed 
in section~\ref{sec:separation}.

{\em Noise and slow drifts}

The presence of slow drifts in the data streams is a well known 
problem to CMB anisotropy measurements.  Long-term instabilities is a 
particularity of the sensitive detector technology used for CMB 
mapping.  Such instabilities originate in a combination of detector 
1/f noise, thermal instabilities, amplifier gain instabilities, 
fluctuations of atmospheric emission (especially for ground-based 
experiments). 
For radio detectors with HEMT amplifiers, the main source of slow 
drifts is 1/f noise. For bolometers, it is more often thermal fluctuations 
of the detector environment that causes trouble.
For instance, a 3\% emissive telescope mirror which temperature 
fluctuates by 0.1~K generates 3~mK fluctuations in detected signals, 
which is mainly in the form of random low-frequency drifts. Removing 
the effects of such slow drifts is known as ``destriping", and will 
be discussed in section~\ref{sec:destriping}.

{\em Imperfections of a real instrument}

Departures from ideality in the instruments must be taken into 
account. This is especially the case for sensitive 
experiments as Planck, for which it is targetted to measure the 
fluctuation power spectrum down to one per cent accuracies per 
single $\ell$ value. 

\begin{figure}[htbp]
  \begin{center}
    \epsfig{file=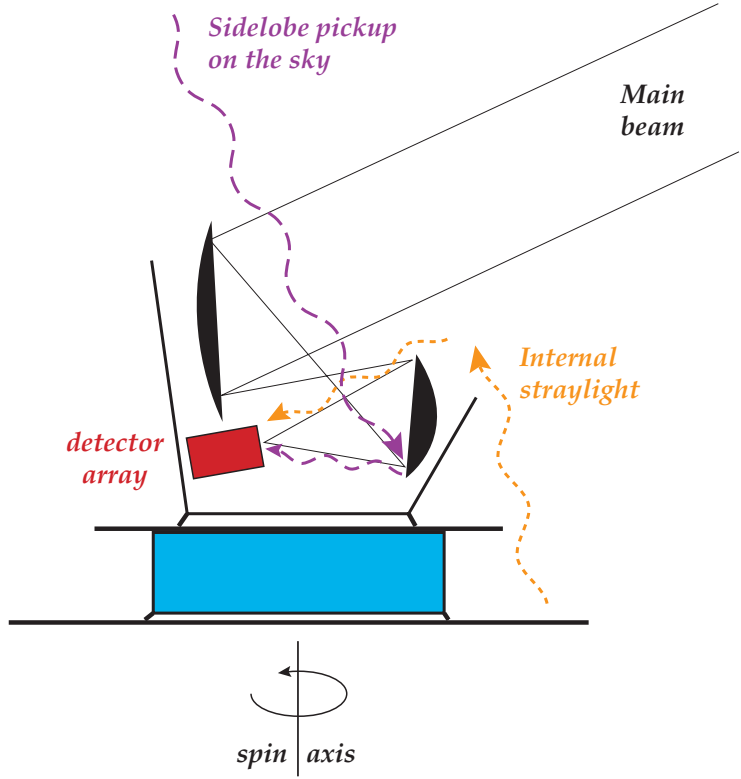,width=.7\textwidth}
    \caption{Illustration of straylight for a satellite as Planck.
    }
    \label{fig:straylight}
  \end{center}
\end{figure}

One critical aspect for Planck is the problem of straylight, which is 
the unwanted radiation received by the instrument detectors.
Straylight can be divided into two broad categories.  The first one is 
straylight originating in the sidelobe pickup of astrophysical 
radiation outside the line of sight defined by the detectors main 
beams.  This problem of ``sidelobe straylight" is well known to 
telecommunication engineers as well as radio-astronomers.  The second, 
more specific to CMB anisotropies (especially at high frequency) is 
that of straylight internal to the instrument, i.e. emission of parts 
of the instrument picked up by the detector.  The emission due to 
fluctuations of the temperature of optical elements, discussed in 
section \ref{sec:separation}, is an example of this ``internal 
straylight". These two aspects of straylight are illustrated in figure
\ref{fig:straylight}.

The estimation of straylight effects for a mission as Planck will be 
discussed in section~\ref{sec:straylight}.

\section{ Separation of astrophysical components } 
\label{sec:separation}

The safest way to avoid contamination of CMB anisotropy observations 
by foregrounds is to make observations in the cleanest regions of the 
sky, where the contribution of the CMB to the sky emission dominates 
by a large amount, so that other components can be neglected.  There 
are a few such regions in the sky, at high galactic latitudes, so this 
has been possible sofar for experiments with little sky coverage.  
Ultimately, it will become necessary to increase sky coverage even for 
small scale experiments, and try to remove foregrounds from the 
measurements, even if they do not impact much the ability of MAP and 
Planck to measure the CMB power spectrum~\cite{knox99}.  For 
astrophysicists anyway, foregrounds are interesting in themselves, as 
they carry information about the properties of interstellar medium or 
about the distribution of astrophysical sources of radiation (and 
hence of matter) in our Universe.

\begin{figure}[htbp]
  \begin{center}
    \epsfig{file=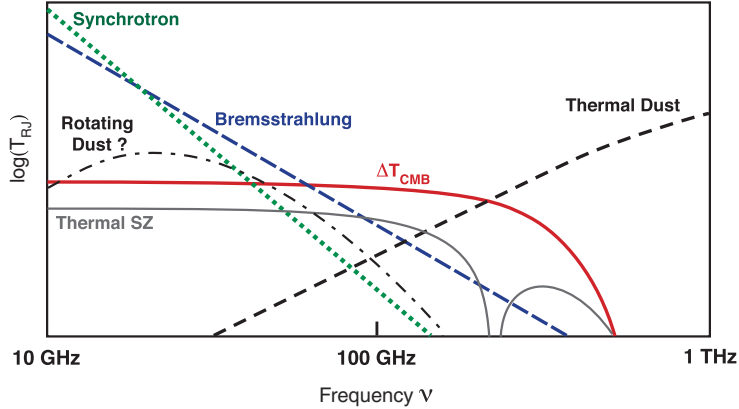,width=.7\textwidth}
    \caption{Foreground emission in the frequency range of interest,
    as compared to CMB anisotropy spectrum. The existence of 
    three major galactic foregrounds, which are thermal dust emission, 
    synchrotron, and bremstrahlung (also called free-free), is well 
    established. There is some evidence towards the existence of 
    microwave emission due to rotating dust. SZ effect generates 
    foreground emission towards clusters of galaxies, and possibly 
    filamentary structure between them. The exact spectra are not 
    known with great precision, which complicates component separation.}
  \end{center}
  \label{fig:foreground-spectra}
\end{figure}

The separation of astrophysical components is possible essentially 
because the components have distinct emission spectra as a function of 
radiation wavelength.  This is illustrated in 
figure~\ref{fig:foreground-spectra}.  If the number of components and 
their respective spectra were perfectly well known, and if 
measurements were noiseless, then multifrequency observations (with at 
least as many observations as there are components) would allow in 
principle the recovery of the emission of each of the component at 
each point.  Equation \ref{eq:simple-model} still holds, but now $t$ 
indexes both time and detector number, and $p$ indexes both time and 
astrophysical component.  The linear system can be large: the total 
number of measurements (i.e. the size of vector $m_{t}$) is $N_{\rm 
detectors} \times N_{\rm samples}$, and the size of vector $T_{p}$ 
(the unknown) is $N_{\rm processes} \times N_{\rm pixels}$.  For 
Planck, the number of detectors is of the order of 100, the number of 
samples is of the order of a billion per timeline, the number of 
pixels of the sky is about one to 10 million, and the number of 
astrophysical processes is at least five or six (3 or 4 galactic 
components, CMB, and SZ - not mentioning complications arising for 
polarisation measurements).

This large system can not be handled by a brute force method.  It can 
be simplified in some cases, if the noise is uncorrelated for 
instance.  Assuming we dispose of clean maps of the sky at various 
frequencies, we can write:

\begin{equation}
    m_{\nu}({\vec x}) = A_{\nu c}T_{c}({\vec x}) + n_{\nu}({\vec x})
    \label{eq:sum-foregrounds}
\end{equation}
where now $\nu$ indexes frequencies of observation, and $c$ indexes 
astrophysical components. $A_{\nu c}$ is the mixing matrix, $m_{\nu}({\vec x})$
the value of the map at frequency $\nu$ in direction $({\vec x})$, and $T_{c}({\vec x})$ the 
emission of component $c$. $n_{\nu}$ is the noise in measurement $m_{\nu}$.

Equation \ref{eq:sum-foregrounds} can be inverted by any of the 
standard methods as the COBE or the Wiener one mentioned in section 
\ref{sec:intro}. This can be done in real space (pixel by pixel), or 
in Fourier space (mode by mode), or on other basis of functions 
(wavelet by wavelet\ldots), depending on the shape and the 
coverage of the maps, the geometry of the observations, and the 
correlation properties of the noise and the components. 

For details about linear methods, we refer the reader to the papers by 
Tegmark~\cite{tegmark97} and Bouchet and Gispert~\cite{fb-rg99}, and for nonlinear 
methods to papers by Hobson et al.~\cite{hjlb98} and Baccigalupi et 
al.~\cite{baccigalupi00}. The Wiener method of Bouchet and Gispert 
has been extended recently to the separation of polarised components by 
Bouchet, Prunet and Sethi~\cite{bps99}.

The trickiest part is getting the right model of the data : assuming 
the right number of components, knowing their spectra (matrix $A_{\nu 
c}$) and knowing the properties of the noise.  This is especially true 
as recent measurements seem to indicate the presence of microwave dust 
emission from rotating grains in addition to the usual thermal 
emission, for instance, which complicate even more the determination 
of the spectrum of the (always sub-dominant, or nearly so) free-free 
emission. Variations of the number of relevant components and of their 
statistical and spectral properties as a function of the region of 
the sky observed complicate the separation of components on full-sky 
maps. Clearly, although some efforts have been made recently in this 
direction, the separation of components by a Wiener method on full sky 
maps is in contradiction with the known properties of foregrounds 
(localised, position-dependent).

Future methods, probably, will have to take explicitely into account 
the effect of uncertainties in the model, including in the error 
estimates the impact of unknown errors in matrix $A$.

\section{ Subtraction of slow drifts } \label{sec:destriping}

Slow drifts are one of the lost fundamental problems we have to face 
for CMB anisotropy measurements. In most cases, this problem 
can be handled satisfactorily independently of the separation of 
astrophysical components. 
This decomposition of the problem of data reduction into subtasks with 
little loss in performance permits to handle very large data sets, as 
discussed by Bond and collaborators~\cite{bcjk99}.

In the simplest model, slow drifts are simply due to $1/f$ noise. The 
total noise (including this low-frequency component) is stationnary, 
its spectrum is known, and the optimal reprojection of the data, 
taking into account the correlation of the noise, can be obtained in 
the {\it COBE} way, as discussed by Tegmark \cite{tegmark2-97}. If the 
spectrum is not known, then one can estimate it from the data 
as in the method of Ferreira and Jaffe \cite{ferreira-jaffe00}. However, 
this last method relies on the assumption that the spectrum of the noise is 
somewhat smooth, as it is binned in ranges of frequencies. This is not 
necessarily the case (actually it is not the case in most of the 
examples I know) so the impact of this uncertainty in the model 
should be evaluated - and, if possible, taken into account for finding 
the best solution and errors.

Another method for slow drifts removal, which does not rely on prior 
knowledge of the noise spectrum, consists in modelling slow drifts as 
a function of a relatively small number of parameters $A_{i}$, to be 
fitted as unknowns of the model. Formally, we write:
\begin{equation}
    m_{t} = A_{tp}T_{p} + f(A_{i}) + n_{t}
\end{equation}

where $f(A_{i})$ is a function slowly varying with time, parametred by 
coefficients $A_{i}$, modelling the effect of the slow drifts, and 
$n_{t}$ can now be assumed to be white.  These coefficients, for instance, 
can be coefficient of the Fourier expansion of the timeline (or 
fractions of it) as in \cite{jd-destripe98}, or coefficients of 
splines used to fit the slow drifts.  The system is then solved for 
$T_{p}$ and $A_{i}$ by minimising the $\chi^2$:

\begin{equation}
   \chi^2 = [m_{t} - A_{tp}T_{p} + f(A_{i})][m_{t} - A_{tp}T_{p} + 
   f(A_{i})]^t
\end{equation}

An adaptation of this method in the context of the measurement of CMB 
polarisation has been implemented on simulated data by Revenu and 
collaborators \cite{revenu00}.

In this method which seems to ``fit the noise out", the fact that the 
noise is essentially low-frequency comes in to allow the low frequency 
drifts to be modelled as a function of a small number of parameters, 
so that the total number of unknown is significantly smaller than the 
number of measurements.  From Shannon sampling theorem, this is 
possible, for instance, if the knee frequency above which slow drifts 
become negligible is significantly smaller than the sampling 
frequency of the data. The slow drifts, then, can be ``sampled" at a 
rate significantly slower than the sampling of the timeline, allowing 
for the number of parameters to be much less than the number of data 
points.

The method, clearly, requires a lot of redundancy in the data, as 
discussed in \cite{jd-moriond2}. These redundancies should also be 
spread out along the timeline so that they allow sampling the slow 
drifts (using parameters $A_{i}$) at a sufficient rate.

\section{The straylight problem} \label{sec:straylight}

One of the most serious problem for most CMB anisotropy experiments, 
and especially for sensitive ones, is the problem of straylight.  
Minimising the total amount of unwanted radiation admitted into the 
detectors is a permanent instrumentalist's nightmare.  The correction 
of straylight effects in the data processing is a complicated problem, 
and is the object of active research.  The first step is the 
estimation of the properties and levels of these effects for a mission 
such as Planck.

\begin{figure}[htbp]
  \begin{center}
    \epsfig{file=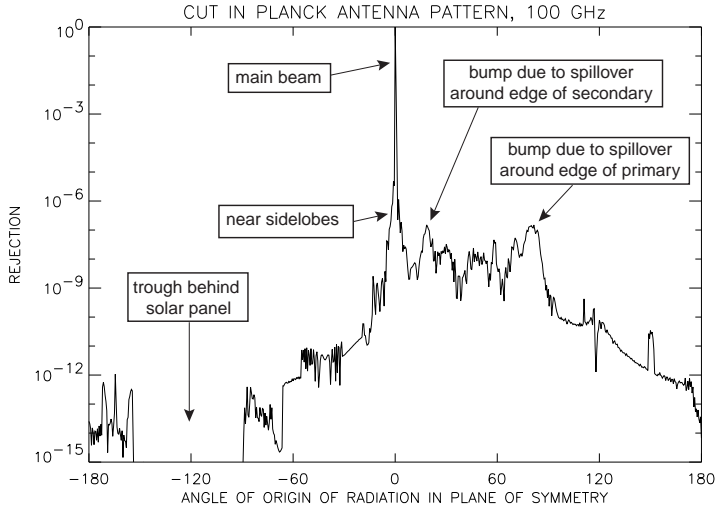,width=.7\textwidth}
    \caption{Cut through a numerical model of the antenna pattern of 
    one Planck 100 GHz channel.}
    \label{fig:lobecut}
  \end{center}
\end{figure}

A model for the Planck antenna pattern has been computed at ESTEC by P. 
de Maagt and collaborators. Figure \ref{fig:lobecut} displays a single
cut of the $4\pi$ antenna pattern (in the plane of symmetry of the 
satellite). 

In terms of integrated emission as seen from solar neighborhood, the 
strongest sources of radiation in the submillimeter sky, relevant for 
the problem of straylight, are the Sun, the planets and the Moon, the 
galactic plane emission, and the CMB dipole.

The properties of the sidelobe pickup depend on the scan strategy. For 
Planck, the satellite is rotated around a spin axis offset by 85 
degrees from the optical axis, so that the main beam corresponding to 
a given detector scans the sky along great circles about 85 degrees 
in diameter. The orientation of the spin axis is re-adjusted every 
hour or so to follow the apparent motion of the Sun in the sky, 
keeping the spin-axis roughly anti-solar. For details on the Planck 
scanning, see~\cite{santanderjd}.

\begin{figure}[htbp]
  \begin{center}
    \epsfig{file=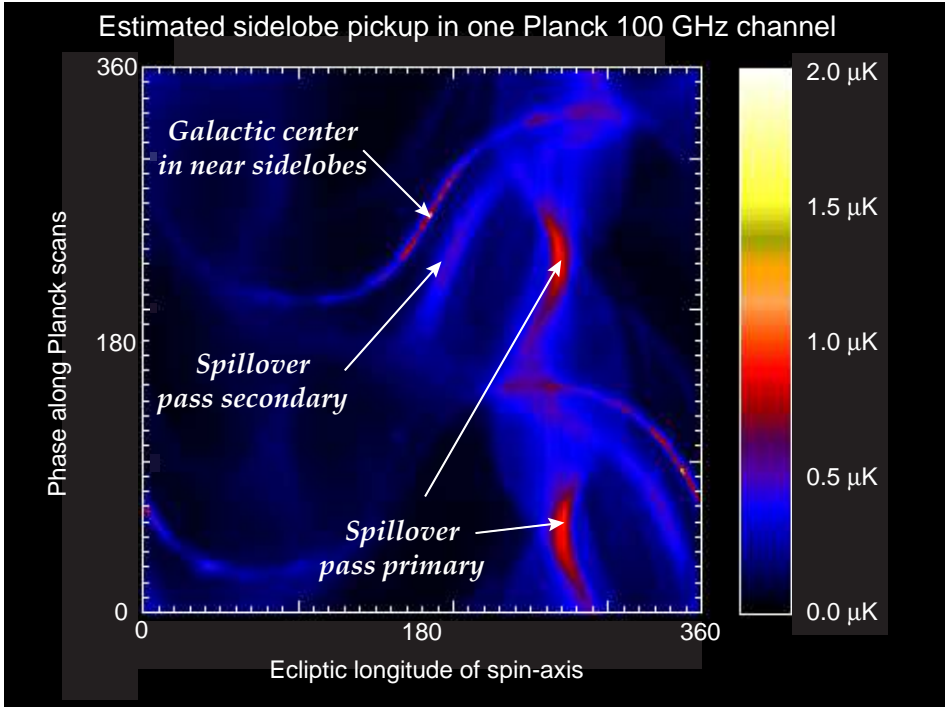,width=.9\textwidth}
    \caption{Sidelobe contribution of the Galactic emission to the 
    Planck signal at 100 GHz. The two-dimensional representation 
    gives the sidelobe signal as a function of the orientation of the 
    spin axis (assumed to be in the ecliptic plane for this 
    calculation) and as a function of the orientation of the 
    satellite around this spin axis. By convention, we fix the origin 
    of each circular scan at the point of closest approach of the main 
    beam to the north ecliptic pole. }
    \label{fig:sidelobe100}
  \end{center}
\end{figure}

Using the Planck 100 GHz antenna pattern of P. de Maagt and a model of the sky 
featuring galactic emission scaled from the DIRBE 240$\mu$m map and 408 
MHz maps \cite{haslam70}, we get estimated straylight signals due to the 
pickup of galactic emission in sidelobes for Planck. The result is
displayed in figure \ref{fig:sidelobe100}. Although the total 
peak-to-peak straylight signal remains within reasonable limits (less 
than 2 $\mu$K), its large-scale properties could generate spurious 
non-gaussianity at low $\ell$, detectable by sensitive non-gaussianity 
tests. It is therefore a potential worry.

\section{Conclusions}

In this paper, I have reviewed the main issues and methods concerning 
the reduction of CMB data. I have emphasised on the importance of 
proper modelling for finding an optimal solution to this problem. The 
main issues for CMB data reduction are foregrounds (astrophysical or 
others), slow drifts of various origins in the timelines, and 
instrumental imperfections as the problem of straylight pickup in the 
detectors. Understanding the issues and the instrumental problems is 
important, as the optimisation of the reduction requires 
optimising the data model at least as much as optimising the 
numerical methods for solving the problem.

Standard linear methods for handling astrophysical foregrounds 
and slow drifts are now well established. These methods could still be 
improved by taking into account explicitely the uncertainty in the 
measurement model (lack of knowledge of the exact number of ``sources" 
for component separation, imprecision in the knowledge of foreground 
spectra and statistical properties, imprecision in the knowledge of 
detector pointings). 

{\em Acknowledgements}

Thanks to Jean-Christophe Hamilton and Alexandre Amblard for their 
critical reading of the original manuscript.

\end{document}